# NATO and Emerging Technologies—The Alliance’s Shifting Approach to Military Innovation

Stephen Herzog

Dominika Kunertova

# NATO AND EMERGING TECHNOLOGIES

## The Alliance's Shifting Approach to Military Innovation

*Stephen Herzog and Dominika Kunertova*

NATO has endured for over seventy-five years, facing the challenges of the Cold War and a difficult transition to counterinsurgency operations after September 11, 2001. Now, the Atlantic Alliance confronts a new set of threats. Revitalized great-power competition and the diffusion of technology undoubtedly will test the adaptability of this thirty-two-nation collective defense organization. Novel technologies hardly are a foreign concept to the world's most powerful military alliance. In its recent history, NATO has helped member countries develop and adopt capabilities ranging from ballistic missile defense to military unmanned aerial vehicles (UAVs), or drones. Many emerging and disruptive technologies (EDTs) of the current era, however, are qualitatively distinct from NATO's previous experiences and therefore pose different challenges.

*Stephen Herzog is a senior researcher at the Center for Security Studies of ETH Zurich, the Swiss Federal Institute of Technology. He is also an associate of Harvard's Belfer Center for Science and International Affairs. In summer 2024, he will become professor of the practice at the Middlebury Institute of International Studies at Monterey's James Martin Center for Nonproliferation Studies. A former U.S. Department of Energy arms control official, he received his PhD from Yale University.*

*Dominika Kunertova is a senior researcher at the Center for Security Studies of ETH Zurich, the Swiss Federal Institute of Technology. She is also a nonresident fellow at the Cornell Brooks School Tech Policy Institute. Previously, she worked at NATO Headquarters in Brussels and NATO Allied Command Transformation in Norfolk. She received her PhD from the Université de Montréal.*

Unlike that of earlier innovations in NATO's portfolio related to improved radar or nuclear weapons, the eventual military utility of nascent EDTs such as artificial intelligence (AI) often is less tangible or apparent. Researchers warn that the performance of AI may soon surpass that of humans in many basic activities such as writing essays and driving vehicles.[1] Recent public fixation with the ChatGPT large language model program points to the vast interest and intrigue surrounding future applications of AI. Meanwhile, quantum computers are beginning to solve complex mathematical problems at speeds far beyond the capacity of humans. Such technologies may be used to decrypt cybersecurity protocols, vastly improve navigation systems, and design and fabricate components for weapons of mass destruction.[2] They also likely will accelerate decision-making speeds and enhance precision-weapon targeting.[3] While militaries have yet to realize the full potential of these technologies, it is not difficult to imagine how EDTs will shape the global strategic environment and future wartime paradigms.[4]

Emerging technologies offer Russia and China tools to contest the liberal international order that the United States and its NATO allies seek to uphold. Their contestation includes Russia's ongoing war against Ukraine and China's plans to project power regionally by enlarging its nuclear arsenal.[5] It is telling that Russian president Vladimir Putin has stated that the country that wins the AI race "will be the ruler of the world."[6] Moscow is pursuing "weaponized AI without any internationally imposed restrictions," with a particular interest in lethal autonomous weapon systems (LAWS).[7] China's leadership has similar views and is attempting to indigenize its semiconductor production—crucial for powering AI software—as the United States tries to restrict Chinese access to foreign technology.[8] Among Beijing's key military objectives are improved intelligence and surveillance, as well as better missile guidance and tracking.[9]

How are these dynamics affecting NATO? The alliance's member countries have heterogeneous levels of investment and interest in EDTs. This complicates efforts to promote alliance-wide innovation, adoption, and standardization for defense purposes. The reality is that the United States spends several times more on emerging technology annually than Europe—collectively—in both the public and private spheres.[10] Washington's initiatives to weaponize EDTs primarily are aimed at countering Beijing, which Capitol Hill and Pentagon decision-makers view as America's foremost military competitor. Many European countries have, however, been hesitant to support the United States in its competition with China and to help secure militarily relevant dual-use technologies.

While NATO increasingly has operated beyond Europe in recent years, Russia's 2022 invasion of Ukraine was a reminder of the alliance's most proximate threat. These changing geopolitical circumstances present an opportunity for the United States to engage the European allies in technological burden sharing. In its plans to deal with an overtly hostile Russia, NATO could serve as a forum to close growing technology and perception gaps between the United States and Europe in various EDT domains.

NATO has shown its potential to be a vehicle for upstream military innovation throughout its history. There are value-added benefits that the alliance could provide to its member countries such as technical expertise sharing, joint industrial development, and organization-wide strategic planning and threat assessment, to name a few. To their credit, NATO leaders have put forward several high-level policy documents to guide these discussions. This is a useful starting point, but the hard work is just beginning. Adapting the alliance to global EDT competition will require transformations in the structure of private-public partnerships and human-capital development. Greater European involvement in these processes would also help sensitize American partners to the importance of EDTs in their downstream military procurement activities.

In this article, we therefore chronicle NATO's shifting approach to military innovation and highlight the challenges lying ahead. We analyze recent key alliance documents on EDTs that remain understudied in the academic and policy literature. Our analysis is complemented by interviews we conducted with high-level officials in Brussels who are involved intimately in the day-to-day work of adapting NATO to the age of EDTs. Our article thus examines NATO's efforts to address new threats, develop partnerships, and foster EDT innovation, adoption, and standardization. Ultimately, we conclude that the Atlantic Alliance has the potential to be a leading forum for military applications of emerging technologies, but getting there would entail overhauling many established bureaucratic practices. Inertia is a daunting, but surmountable, hurdle.

## GREAT-POWER COMPETITION, ALLIANCE POLITICS, AND EMERGING TECHNOLOGIES COLLIDE

NATO's security context changed dramatically across 2021 and 2022. The last coalition troops left Afghanistan in August 2021, ending nearly two decades of out-of-area military operations, the largest undertaken by NATO countries. This signaled a return to focusing on the territorial defense of Europe as great-power competition unfolded. The logic underlying this decision seemed vindicated when the Kremlin began its unprovoked, full-scale invasion to conquer neighboring Ukraine on 24 February 2022. Putin's war marked a new period of economic and political closeness between Moscow and Beijing, key dissenters from the U.S.-backed liberal international order. China has also continued its quest for regional hegemony in East Asia, upping its threats to Taiwan's sovereignty and making efforts to "modernize, expand, and diversify its nuclear arsenal."[11] It is no wonder there is increasing interest in preparing the alliance for great-power competition.[12]

The new strategic concept adopted by NATO at the June 2022 Madrid Summit directly addresses these new security realities.[13] Three central points are notable for great-power competition and EDTs. First, the concept is unambiguous on Russia: "The Russian Federation is the most significant and direct threat to Allies' security and to peace and stability in the Euro-Atlantic area."[14] Second, for the first time, the Atlantic Alliance specifically highlighted China as a threat. The reasons for this include its global ambitions, partnership with Russia, and efforts "to control key technological and industrial sectors, critical infrastructure, and strategic materials and supply chains."[15] Third, the strategic concept warns of adversary nations' use of EDTs. It reflects a recognition that NATO needs to "promote innovation and increase our investments in emerging and disruptive technologies to retain our interoperability and military edge."[16]

Russia presents an immediate threat on NATO's doorstep and leverages EDTs in pursuit of its political and military objectives. When announcing his invasion of Ukraine, Putin issued nuclear threats to deter NATO's intervention in the conflict.[17] Moscow also suspended its participation in the New Strategic Arms Reduction Treaty with Washington and used nuclear-capable (but conventionally armed) and purportedly hypersonic Kinzhal missiles against Ukraine. Furthermore, the Duma withdrew its ratification of the Comprehensive Nuclear-Test-Ban Treaty. The Russian military actively is leveraging machine learning and autonomous systems in its "nuclear command, control, and communications . . . and warfighting capabilities."[18] These developments improve the precision of Russia's nuclear targeting and the ability of its missiles—many aimed at North American and European capitals—to evade air defenses.[19] Meanwhile, Russia has unleashed drone warfare on Ukraine and its population using both indigenous systems and units procured from China and Iran.[20] Some drone types, like the Kalashnikov KUB-BLA UAV, depend on AI algorithms to autonomously identify targets.[21] While Russia's investment in EDTs lags behind the United States and China, it is apparent that the Kremlin has an interest in coupling AI with its existing military capabilities, such as drones, where the "Russians are well aware that swarm technology, powered by artificial intelligence, is seen as a significant force multiplier."[22]

The longer-term challenge to the liberal international order, however, emanates from Beijing, not Moscow. Chinese posturing is not limited to expanding its nuclear arsenal or military drills to intimidate Taiwan and its Western backers. China's rush to secure semiconductors and its sprawling industrial espionage programs aim to close military capability gaps with the United States.[23] For several years now, Chinese defense documents have identified the United States as the country's main adversary and have pointed to military technology as a potential equalizer.[24] Under the Chinese military-civil fusion national strategy, there is little distinction between private and government entities; all EDT investment and development can be used to enhance and project military power. The resulting range of initiatives is vast: "intelligent and autonomous unmanned systems; AI-enabled data fusion, information processing, and intelligence analysis; wargaming, simulation, and training; defense, offense, and command in information warfare; and intelligent support to command decision-making."[25]

Intra-alliance divisions present an obstacle to protecting NATO from potential Russian and Chinese aggression with EDTs. Regarding Russian hostilities, the June 2022 Madrid Summit Declaration was unequivocal: "NATO will continue to protect our populations and defend every inch of Allied territory at all times."[26] But trade in dual-use technologies, especially those with unclear end states, is more divisive. Because of geographic proximity, Europeans long have had comparatively more economic integration with Russia than their American and Canadian allies have had with Moscow. The European Union (EU) was Russia's largest trading partner, and Russia was the EU's fifth-largest, before the war in Ukraine, with considerable trade in energy and machinery.[27] Several European states, such as Germany, initially encountered resistance to sanctions from homegrown firms engaged in information technology transactions with Russia.

There is even less alliance consensus on China, despite the rhetoric of NATO's new strategic concept. Part of this has to do with Europe's distance from China, and part stems from trade ties. The EU's March 2022 Strategic Compass document demonstrates the complexity of relations with China: "China is a partner for cooperation, an economic competitor, and a systemic rival."[28] On the one hand, most of NATO's eastern European allies are squarely with the United States and United Kingdom in viewing China as a competitor and rival. On the other hand, Paris and Berlin have deep economic partnerships with Beijing. Nowhere was this clearer than in French president Emmanuel Macron's controversial remarks to journalists after visiting China in April 2023. Macron spoke of preventing Europe from becoming "America's followers" and not getting "caught up in crises that are not ours," such as the Taiwan standoff.[29] His statements were praised by leaders in Beijing, who encourage European strategic autonomy to erode NATO cohesion and prevent broader Western involvement in the Indo-Pacific.[30] Another example of division is over whether to allow Chinese telecommunications giant Huawei to provide infrastructure for 5G mobile Internet networks in Europe, which will eventually be replaced with AI-enabled 6G.[31] Some states have signed contracts with Huawei, and others have banned it—fearing the introduction of "choke points of vulnerability" given China's military-civil fusion strategy.[32]

Great-power competition in the realm of EDTs creates new mission space for NATO as a collective defense organization. The alliance needs to protect its intellectual property and critical resources while also finding ways to inspire military innovations to be adopted and standardized. This is no easy set of tasks; it will require significant restructuring of some of the ways the alliance has pursued business for generations.

The traditional Western military-industrial complex may excel at designing main battle tanks, field artillery, and jet fighter aircraft, but it is not optimized for EDTs. Transatlantic defense industry stalwarts such as Airbus, BAE Systems, Boeing, Lockheed Martin, and Raytheon have rushed to invest in AI, machine learning, and quantum technologies. Yet this has often been done in partnership with information technology companies that have considerable processing power and expertise, because large defense contractors are not attracting "whiz kid" coders like information technology firms. For this reason, one senior NATO official told us that the tightly controlled military-industrial complex was being replaced with something new.[33] These thoughts reflect the idea that the top-down approach to procurement requires adaptation. Military innovation in EDT domains means securing access to the firms with the right talent, corporate culture, and environment to facilitate creativity.

One proverbial elephant in the room is that young professionals who flock to Silicon Valley and other innovation hubs usually are not excited to work on military projects. In October 2022, a group of robotics manufacturers, including industry leader Boston Dynamics, pledged not to weaponize their products for LAWS or other purposes.[34] This trend has led to numerous warnings that the West will eventually fall behind Russia and China in the EDT race. A *Financial Times* commentary calling for European defense innovation was particularly blunt, stating, "Democracy won't defend itself with the next grocery-delivery app."[35] The stigma effect is all too real for NATO and the Boston Dynamics pledge spurred officials to think about how to improve their reputation in the private sector.[36] The same concern applies to the alliance's own personnel recruiting and that of member countries' ministries of defense. These bodies need tech-savvy staffers to assess military needs and shape procurement decisions.

European states have also shown markedly less interest than the United States in EDT-driven military innovation. Defense spending in Europe since the 2008 financial crisis is both uncoordinated and reduced from previous levels.[37] As a result, Washington has taken the lead in directing the alliance's path forward on emerging technologies. Although recent strategic documents suggest that China poses a threat to NATO, there are fractures in the alliance over whether to back the U.S. strategic pivot to Asia. Much of U.S. defense planning in the EDT realm is directed at containing and countering China. European investment in military innovation and technological burden sharing would thus serve two purposes: responding to Russian EDT efforts, and educating Europe's leaders about the dangers posed by commercial relationships with certain Chinese technology firms.

## NATO ADAPTS TO THE NEW TECHNOLOGICAL ENVIRONMENT

It would not be fair to say that the Atlantic Alliance has stood by idly as threats to the European continent have evolved. Though its member countries have their own militaries, there is a role for NATO in the context of multipolar competition with strong technological dimensions. It was always difficult to forge a strategic partnership with Moscow. In the case of China, the alliance is entering uncharted waters without an organization-wide playbook. EDT-driven innovation is fast becoming one of the main elements of NATO's ongoing adaptation to Russian and Chinese subversion of the rules-based international order.[38] The alliance's response is organized around three principal motivations.

First, new technologies can compensate for negative demographic trends.[39] While the United States has a relatively young and growing population, declining European fertility rates and aging patterns pose challenges to fielding operational forces.[40] The force-multiplying effect of EDTs can aid alliance troops. These technologies can convey advantages of speed, precision, and autonomy to keep soldiers out of harm's way, improving the efficiency of military missions. Energy-efficient emerging technologies also may reduce greenhouse gas emissions of military deployments as the security implications of climate change loom large.[41]

Second, steering the dynamics of defense innovation will allow the allies to set EDT safety standards and ethical codes. Analysts fear that if autocracies are the first movers in domains such as AI, the risks of accidents and unethical use will increase dramatically.[42] By contrast, the view among experts in Brussels is that investment in military innovation by NATO member countries will shape the responsible use of technologies toward liberal democratic values and human rights.[43] The alliance is therefore striving to formulate normative principles for EDT governance.

And third, technological leadership has acquired geostrategic significance. Such considerations feature prominently in U.S., Chinese, and Russian discourses alike.[44] NATO leaders publicly declared in 2021 that the alliance needed to master EDTs to avoid vulnerabilities and that "malicious use" undermines their security.[45] This empowered the statement that increased EDT innovation would "help to ensure, individually and collectively, our technological edge now and in the future."[46] The notion of "technological edge" alongside allied force interoperability has always been central to NATO's defense and deterrence posture.

NATO's history and structure make it well positioned to adapt to EDTs. It remains the only organization providing a daily forum for European and North American political leaders to coordinate on security. In addition to its core principle of collective defense, the alliance reduces transaction costs of cooperation and improves information sharing.[47] NATO is reputed for setting military standards in technical and operational areas to maintain allied force interoperability. Beneath these strategic concept–level tasks, NATO institutionalizes iterated civil and military cooperation among member countries. Its lesser-known contribution to collective defense lies in developing and adopting new technologies. NATO creates a framework for aggregating its members' military technology developments.[48] In the past, the alliance spearheaded multinational development of precision-guided munitions and helicopters, to name a few initiatives, and the safety of an allied defense marketplace is why the American F-16 Fighting Falcon fighter and German Leopard 2 tank are used so widely by European militaries. In the future, the transatlantic defense economy inevitably will feature more development and trade in computer software and dual-use technologies.[49]

Put simply, the diffusion of technology can spur military innovation. According to the literature, innovation usually encompasses changes in how troops function in the field, with improved military effectiveness.[50] Innovation may range from adaptations of existing technology or tactics to major changes in the conduct of warfare that alter the core competencies of organizations and soldiers alike.[51] In line with these definitions, mature EDTs are expected to offer militaries improved communications, navigation, targeting, and more. But upstream innovation also requires an actor's downstream capacity to adopt and ultimately put new technologies to use.[52] Indeed, the development and fielding of a new military system can be hindered by resource constraints and integrative deficiencies such as inflexible command, obsolete doctrine, and insufficient training. NATO's efforts to ensure interoperability have been pivotal to overcoming these barriers in the past.

Research suggests that a country's participation in NATO and the external threats presented by Russia and China should encourage its military innovation.[53] For smaller states in particular, the alliance will play a large role in these activities as strategic competition increasingly involves EDTs.[54] NATO's strides to innovate in several converging domains may help its smaller members close capability gaps and find their technology niche. Yet studies on alliance management are mostly silent on industrial policy.[55] NATO has in recent years, however, become more involved in joint procurement and multinational capability development projects. Beyond providing defense expertise for members' innovation efforts, NATO has endeavored to build strong ties among political leaders, military authorities, and armaments manufacturers. There is also a growing recognition that dialogue on EDTs and forums for funding technology will require further involvement of private-sector and academic experts.[56]

This approach parts ways with decades of NATO practice regarding military innovation. The alliance initially engaged with technology to produce platforms for specific applications. Its Cold War innovation strategy was to develop weapon systems that were technically superior to those of the Soviets—largely facilitated by the U.S. defense industry.[57] After the collapse of the Soviet Union, NATO technology discussions lost their competitive narrative. Instead, the allies primarily focused on defensive applications. The predominant threats of the 1990s and early 2000s were not, after all, great powers. The fight against terrorism included concerns about the acquisition of weapons of mass destruction as well as the use of improvised explosive devices. Leaders also worried about new vulnerabilities for the European continent as "rogue states," especially Iran and North Korea, sought ballistic and cruise missiles.

Until the early 2010s, there even was an appetite within NATO for technical cooperation with Russia. One potential area of collaboration considered was missile defense to defend against potential attacks from Iran and North Korea. After Russia annexed Crimea in 2014, this dialogue collapsed.[58] Aside from treaty-mandated arms control verification between Washington and Moscow, most science and technology cooperation stalled. Dialogue on military technology issues with Russia within NATO's Euro-Atlantic Partnership became nearly impossible. When NATO-Russia ties looked like they could not get worse, Putin launched his full-scale invasion of Ukraine in 2022, and the remaining vestiges of technology cooperation for arms control, business, and scientific exploration came to an abrupt end.

There is a common thread in NATO's history of technological competition and cooperation with Russia—the alliance's model of military innovation focused on specific platforms, capabilities, and partnerships with the traditional military-industrial complex. The changing strategic context discussed in the last section demonstrates why NATO, as an organization, realized this approach was no longer viable in the 2020s.

Now, the innovation model is technology driven and relies on the private sector.[59] This is what officials believe is necessary for great-power competition with Russia and China. During the Cold War, militaries exported technologies into the commercial sector. In turn, commercial companies did not shy away from defense contracts, then regarded as more lucrative than consumer markets. These contracts also presented future possible sales of technology derivatives with civilian and consumer applications. Today's innovation process is different, as it is occurring in the laboratories of large technology firms and in start-up garages. Meanwhile, defense companies are spending less on in-house research and development.[60] The result inverts the innovation pyramid, and militaries are importing technology primarily developed for the commercial sector. Even the advanced weapons and aerospace technologies that are still safely the province of established defense contractors are quickly being integrated with EDTs.

Military and political leaders at NATO Headquarters in Brussels seem to be aware of problems created by global strategic competition and the commercial roots of new military technologies. For instance, nondemocratic rivals may strike deals with Western firms so that they can eventually weaponize technologies that have indeterminate outputs or ambiguous military applications. In many cases, young software developers and their managers may be unaware of the implications of such transactions. The armed forces of NATO members always have attempted to leverage technological advances and their industrial bases. But the promises and perils of EDTs are relatively new to the alliance's high-level political documents and strategic communications, which indicates that policy makers are just beginning to pay attention to these technologies. The rise of EDTs entails imperatives for competition alongside the protection of certain human-capital resources and dual-use commercial sectors.[61]

In reality, despite this lack of high-level attention, NATO has been monitoring EDTs for over a decade, principally to counter asymmetric adversaries.[62] New challenges related to technology innovation arrived in the 2010s, including the digitization of NATO's command structure and the development of a joint intelligence, surveillance, and reconnaissance architecture. Over the past decade, NATO established its own organizational fleet of ground surveillance UAVs and designated cyber and space as operational domains.[63]

The technologies themselves, however, did not receive high-level political attention until several years later. NATO's 2010 strategic concept contained a single passing reference to "ensur[ing] that the Alliance is at the front edge in assessing the security impact of emerging technologies, and that military planning takes the potential threats into account."[64] The document did not highlight specific technologies or make recommendations about NATO's own military innovation. By 2022, things had changed monumentally. The strategic concept adopted in June 2022 explicitly highlights EDTs as a key arena for NATO, given global competition. Furthermore, it acknowledges technological primacy as a condition for battlefield success and commits its member countries to innovating and investing in EDTs.[65]

The breadth and depth of the 2022 strategic concept's engagement with emerging technologies should not have surprised close followers of NATO.[66] The alliance's interest in EDTs became more concrete in 2019 with the approval of a road map to structure its work across seven key technology areas (as described in the road map): AI, autonomy, biotechnology, data, hypersonic, quantum, and space. NATO officials also have identified electronics and electromagnetics, energy and propulsion, information and communication systems, and novel materials and technology as worthy of increased attention.[67] The 2019 road map represented the first effort to engage directly with technologies in terms of innovation rather than assessment and defense, which likely explains the (perhaps overly) broad nature of some of the designated technology areas.

The road map paved the way for further attention to EDTs. Technological innovation was no longer just the purview of working-level specialists within NATO; it had gained high-level visibility in the North Atlantic Council and NATO senior committees, which approved a comprehensive EDT strategy in February 2021.[68] The strategy laid out political goals and specified a two-pillar approach. First, NATO member countries pledged to foster innovation by supporting EDT research and development. Second, the alliance warned of the need for strong national export controls and soft regulatory norms to protect innovators and technology against misuse.[69]

At the June 2021 Brussels Summit, NATO introduced steps to adjust its top-down model of military innovation. Over the past seventy-five years, the alliance has developed close working relations with the defense industry, evidenced by its high-profile NATO Industry Forum and the NATO Industry Advisory Group. These venues do not allow for truly interactive engagement with commercial firms, however. But in Brussels, national leaders sought to reverse this trend by creating conditions for bottom-up developments on EDTs within the NATO technology ecosystem. The centerpiece was a new civil-military technology innovation partnership called the Defense Innovation Accelerator for the North Atlantic (DIANA). Additionally, twenty-three member countries agreed to establish a €1 billion NATO Innovation Fund to support start-ups working on dual-use EDTs over a fifteen-year period.[70]

DIANA aims to attract start-up, software-oriented firms whose programmers have no prior experience working with the defense sector. This accelerator will let private companies compete to provide innovative solutions to problems faced by NATO. In exchange, the alliance offers financial grants, mentorship in working on defense projects, and access to end users on both sides of the Atlantic. DIANA is developing into an EDT industry complex within NATO. It has offices in Canada and the United Kingdom, a regional hub in Estonia, and over one hundred accredited testing centers for use by innovators.[71] Some NATO officials hope DIANA will serve as a national blueprint for member countries.[72]

In the long run, NATO must address a bias prevalent among civilian researchers that defense research is unethical. Because NATO has faced difficulty establishing credibility in the eyes of the private sector in the past, the alliance's evolving approach to military innovation involves discussions of responsible-use principles based on liberal democratic values and respect for human rights. Soft-norm regulatory frameworks such as these allow military actors to signal values to private-sector companies. The goal is to convey to private firms that NATO is an ethical and trustworthy organization and to show the world it intends to shape responsible technology governance.[73] As of April 2024, NATO has published four informal codes for propagating soft norms. They pertain to the responsible uses of AI, autonomous systems, quantum technologies, and biotechnologies and human enhancement. The organization also established its Data and Artificial Intelligence Review Board to begin working on a "NATO Responsible Artificial Intelligence certification standard."[74]

Officials understand NATO's EDT strategy, codes of conduct, and investments in innovation as vital elements in maintaining collective defense.[75] Involvement in the emerging technology race has become part and parcel of NATO's core mission. Given our interviews with senior-level officials working on EDTs and our close reading of strategy documents, we identify four functions that enable NATO's changing way of promoting military innovation in EDTs:

1. *Generating* in-house expertise for identifying and developing EDTs as well as addressing threats posed by technological developments
2. *Enabling* technology adoption and integration via standardization and individual or joint procurement support
3. *Networking* with—and investing in—innovators to secure access to the private sector while safeguarding sensitive technologies from adversarial influences
4. *Regulating* military applications of technology through establishing value-based principles of responsible use to shape technology governance

Crucially, NATO provides the necessary political and strategic context for its member countries' EDT engagement. The alliance is adapting to provide resources for understanding the technological availability, technical feasibility, and military utility of new systems. It also fosters cooperation among scientific and technical communities across NATO countries to build trust, share knowledge, and develop expertise they can take home. Coordinated testing and operationalization of EDTs have also produced new NATO standardization agreements on emerging technologies in areas such as aerial and ground robotics.[76] The intent is to further enable the adoption of EDTs at the national level.

Framed in terms of either governance or competition, technology now shapes global politics. NATO is developing tools not only to navigate the era of EDTs but also to actively influence technological progress. Adapting an alliance created in the aftermath of the Second World War to define future capabilities based on digital technologies is no easy feat. However, the more serious challenges lie in maintaining NATO's technological edge and interoperability among member countries. Uneven development could undermine cohesion and the allies' ability to work together.

## EMERGING BLIND SPOTS

This new era that combines great-power competition and EDTs is very different from previous periods of NATO military innovation. Emerging technologies are *capability agnostic*, *privatized*, and *multidomain*. In this section, we show how these dynamics present a series of challenges to the Atlantic Alliance in developing and adopting EDTs.

The technology landscape of the past revolved around discrete platforms to provide specific capabilities, fabricated by the traditional military-industrial complex. But extracting benefits from today's technology-driven innovation requires well-defined requirements and informed policy guidance. Stated differently, the armed forces are buying solutions, not technology.[77] Meanwhile, the power to shape and control military innovation has moved to the private sector, requiring a new interactive cycle beyond legacy procurement processes and traditional defense contractors. NATO has to inform commercial innovators about military needs and then bring awareness back to the armed forces. The alliance cannot leverage advantages in emerging technologies without cultivating relationships with new defense and nondefense partners from the private sector.

The military revolution just over the horizon may bring not one but potentially many "Sputnik moments" across technology domains. The Soviet Union's launch of the first satellite into space in 1957 shook NATO's presumption of possessing a qualitative technological advantage over its adversary. The launch also carried military ramifications; allies' capitals would soon become vulnerable to Soviet ballistic missile developments. The alliance reacted by interpreting article 2 of the North Atlantic Treaty pertaining to "economic collaboration" to apply to peaceful science and technology fields.[78] In 1958, the allies created the NATO Science Committee, which eventually led to today's Science for Peace and Security Program. Global media resurrected the Sputnik metaphor most recently in 2021 in response to China's test of an orbital hypersonic glider.[79] Avoiding Sputnik moments can be difficult when the fast pace of innovation sharply contrasts with the slow pace of military procurement. The difference in the timescale of innovation cycles between the commercial sector (months) and the military (years) can hamper countries' ability to harness new technology. Hence, recent scholarship has concluded that NATO needs a "common strategic culture of innovation."[80]

NATO's EDT strategy discussed in the previous section does appear—at least rhetorically—to address these shifting dynamics. The practical implementation of these policies is not without challenges. From our research, we can identify five:

1. The emerging state of technologies makes concrete military deliverables a moving target.
2. Low rates of technology literacy create space for politicizing EDTs.
3. Staffers with technical expertise are difficult for military organizations to recruit and retain.
4. New technology-driven innovation may clash with preexisting capabilities and industrial partnerships.
5. Monitoring compliance with certifications and codes of conduct relies on delegation to national authorities.

First, the characteristics of emerging technologies are still evolving, and military outputs are not always immediately apparent. The resultant military innovation timeline becomes cyclical rather than linear.[81] The alliance may find it difficult to specify requirements for defense without limiting a technology's potential, something that may prove unpopular with innovators. An inability to visualize fully the component or system at the end of the pipeline can complicate assessment and peer-review processes. Personnel working on DIANA and NATO Innovation Fund activities may be pressed to create evaluation metrics that resemble more closely those of civilian scientific bodies than those of a military organization. This risks some technologies evolving in such a way that they may not ultimately become adoptable by the armed forces.

Budgetary constraints for funding capability-agnostic technologies present another predictable limitation. DIANA is known internally as NATO's analogue of the U.S. Defense Advanced Research Projects Agency (DARPA) because of its focus on military applications of EDTs. While DARPA has an annual budget of over U.S.$4 billion, DIANA's budget is only €50 million.[82] This almost symbolic financial incentive may help with idea incubation but likely will be insufficient to fund a start-up whose existence depends on a sole product. Ideally, national authorities would provide further financial assistance to promising entrepreneurs. Similarly, the NATO Innovation Fund depends on the €1 billion pledge over fifteen years made by twenty-three member countries. Though funds are limited, NATO innovation initiatives already have competition. In May 2022, the European Defence Agency inaugurated its Hub for EU Defence Innovation to support cooperation and manage networks of organizations and researchers.[83] NATO and the EU need to leverage their respective institutional strengths and memberships better to avoid duplicating efforts. Intelligence sharing among the members and between both organizations—a historical bone of contention due in large part to Turkey-Cyprus tensions—would help protect the European industrial base against exploitation by rival actors.

Second, improving technological literacy within the alliance could help to avoid the hype and politicization of EDTs. At first glance, NATO is prepared to lead discussions about EDTs across its structures and member countries. It is already assisting countries in this regard and supporting national research and development strategies. However, it is fast becoming clear that NATO needs "translators" to facilitate knowledge sharing between political leaders and engineers, and between military technicians and soldiers.[84] These groups have different day-to-day responsibilities, timelines, and interactions with technology. When Sputnik moments occur, EDT developments need to be placed in an appropriate technical context before leaders react and speak to the media.

As an alliance, an issue for NATO is how to manage its political leaders' expectations regarding EDTs. Preconceived notions of ever-more-sophisticated emerging technologies can lead to turning a blind eye to more-rudimentary ones.

The war in Ukraine is making military applications of EDTs—such as off-the-shelf commercial drones—less abstract and more basic than media speculation of technical marvels suggests.[85]

NATO has the capacity to produce its own technology foresight. Its Science and Technology Organization (STO) provides leaders with advice on notable technological areas and assesses the impact of EDTs on defense and security. The STO has provided an in-depth examination of the transformative and revolutionary potential of EDTs over the next two decades.[86] The organization also raises awareness that individual types of EDTs are unlikely to have impacts on their own. Instead, their major effects on military competition and international order will come in clusters such as data-AI-autonomy or space–hypersonic technology–materials.

The STO paradoxically was not involved in drafting NATO's 2019 EDT road map, which initiated the alliance's posture on emerging technologies. Without the participation of the STO, NATO's discussions lacked real technical expertise.[87] Items that are not technologies per se were included on the list of relevant EDTs, such as data, space, and autonomy. Yet in the past decade, nearly half the STO studies dealt with technologies now labeled EDTs, including drones and hypersonic capabilities.[88] The alliance's defense planning addressed both. NATO thus is rebranding its existing work and taking account of where its expertise lies. The 2021 EDT strategy aims to streamline these processes.

The general lack of European interest and investment in military innovation is producing an intra-alliance technology gap between the United States and Europe. We have discussed the political implications of heterogeneous EDT investment, but there are operational effects as well. Without revitalized European spending on military innovation, the gap will grow as the United States pursues its defense-innovation offset strategy, popularly known as the "third offset."[89] The situation risks making practical transatlantic cooperation between allied forces more difficult, if not impossible.[90] NATO initiatives to foster EDT innovation unintentionally could hurt interoperability if other members do not close this technological gap between the United States and the rest of the alliance. The alliance may also serve as a forum to coordinate positions when European countries are struggling to agree on a coherent approach toward China. NATO could offer its members strong leadership on EDTs and a reliable market for technology trade to present a credible alternative to Beijing. The latest U.S. National Defense Science and Technology Strategy places explicit emphasis on collaborating with allies and developing interoperable technology.[91] Whether this will appeal as a template for the alliance remains to be seen.

Calls for greater technological burden sharing have been part of NATO dialogue for decades. Military technology cooperation among the allies has always been a key element of the alliance's fabric, but "achieving rationalization, standardization, and interoperability of Allied weapons has proved to be an elusive goal."[92] In the strategic competition for EDT supremacy between the United States and China, a new layer to the transatlantic bargain is beginning to include technology decoupling from China (or the more politic "de-risking").[93] The balancing act between U.S. technological leadership and recent calls for European technology sovereignty may thus harm Atlantic Alliance cohesion.

Third, NATO and its member countries face shortages of bona fide technical experts in EDT domains. Establishing and implementing the 2021 EDT strategy thus became a collective learning exercise involving roughly 25 percent of NATO personnel.[94] These discussions provided on-the-job learning, which was especially important for delegations from small countries without the capacity to do policy making on EDTs. However, NATO has cannibalized its staff owing to its EDT expertise shortage, taking key experts away from other areas of the organization. The situation raises the question: Who innovates the innovator? While harnessing innovation is never straightforward, militaries are usually resistant to the new and prone to preserve old ideas and practices for the sake of stability.[95] But in the case of EDTs, military innovation necessitates institutional change.

The alliance has tried to solve its human-capital problem with a top-down approach, one that does not yet appear to be working successfully. NATO leadership has issued several high-level political statements and set up senior internal (NATO Innovation Board) and external boards (NATO Advisory Group on EDTs, NATO 2030 Reflection Group). The objective is to adapt the alliance to technology-driven military innovation dynamics and draw in private-sector partners. At present, the initiatives may be too far removed from the daily activities of NATO's staff. For the alliance to innovate itself, forming external advisory groups may not be enough to nurture changes in established practices and culture.[96] Attracting human talent with the technical skills to assess EDTs, related global events, and military needs is central to NATO's ability to inspire technological innovation and adoption.

Fourth, institutional path dependence may hinder EDT adoption. NATO has designed DIANA to obtain new technology know-how quickly, but downstream integration of technology into its command structure and various members' national militaries is another task. This challenge is multifaceted and involves overcoming reliance on legacy systems, reforming lengthy acquisition processes, deconflicting new initiatives with preexisting industrial partnerships, integrating new technologies into extant systems and mind-sets, and restructuring armed forces. Oftentimes making a new technology work for the military is more difficult than the dynamics of the innovation pipeline itself. Deeper engagement with the private sector would be a start in changing the alliance's approach to adopting novel technology. Structural changes to the organization such as promoting budgetary flexibility and hiring technical acquisition officers could also prove useful for military innovation.[97]

The technology problem is systemic and recurrent. In the 1980s, European countries encountered procurement difficulties and developmental delays with then-emerging technologies, including sensors for tactical reconnaissance capabilities, and microchips and microcomputers for missile guidance and fire-control systems.[98] These issues were compounded by standardization procedures and incompatibilities between U.S. and European defense systems.[99] Matching existing standards and system interoperability requirements with EDTs entails considerable intellectual labor. This includes both greater vertical technical-tactical interoperability and the integration of EDTs into practical training and exercises. Militaries cannot incorporate modern technologies into antiquated planning and mind-sets effectively. Importantly, technology hype and dogma can present setbacks to NATO if the alliance is not prepared to balance innovation with appropriate feasibility criteria. Decision makers typically understand innovation in terms of improving military combat readiness and effectiveness. But innovation also can hinder the achievement of both battlefield and political objectives, especially alongside growing security commitments and shrinking resources.[100]

Fifth, NATO is an intergovernmental organization with neither the capacity nor mandate to monitor technology adoption and enforce compliance with its standards. Stated differently, upstream innovation and downstream adoption processes are distinct. Take the NATO AI strategy, for example.[101] The allies' defense ministers collectively endorsed it in 2021 and the strategy is now backed by a NATO Responsible AI User certification. The success of NATO policy making on AI, however, depends entirely on national implementation and the efforts of member country authorities. This is true in terms of the strategy's guidance on informed decision-making and on developing interoperable systems. That said, policy and industrial stakeholders may hope to garner NATO support for national innovation efforts, which could encourage compliance.[102]

Publishing ethical codes for soft norms in technology governance that the organization cannot enforce could have mixed effects. As we discussed above, doing so signals to the international community and prospective private-sector partners. At the same time, being among the first actors to publish such principles may cost NATO in terms of strategic advantage and military effectiveness. Russia and China are unlikely to have such constraints. And if some member countries opt to ignore NATO's purported normative values, it hurts the credibility of the organization as a whole.

## IS NATO READY FOR THE LOOMING MILITARY REVOLUTION?

The Atlantic Alliance faces significant challenges at the intersection of great-power competition with Russia and China and emerging technologies. NATO is a seventy-five-year-old military organization that needs a push from above to change. Our research suggests that the evolving strategic context and prevailing beliefs in technological edge and solutionism prompted decision makers to address the long process of adapting to the new era of defense innovation. NATO leadership has created a new set of tools to encourage bottom-up innovation to find its way into hierarchical structures long veiled by a classified military culture. If this new approach succeeds, it will be because the alliance cultivated relationships with private industry, overcame hurdles to adopting new systems, and set the tone for dual-use technology regulation.

Our contribution to understanding military innovation that is capability agnostic, privatized, and multidomain is both conceptual and empirical, illustrating how military innovation and technology may come together within contemporary security alliances. Military innovation in the age of EDTs requires institutional change. The main determinants appear to be threefold. First, EDTs necessitate civil-military innovation involving both political leaders and commercial innovators. Second, innovation requires strategic and cultural shifts within organizations due to the security environment and the pace and scope of technological advances. Third, NATO's approach to technology-driven military innovation can be explained through temporal sequencing. The alliance's leadership took steps to create conditions for bottom-up innovation with new infrastructure and financial incentives. The intent is for the private sector to innovate solutions to NATO-defined problems. This interactive and inclusive model parts ways with the rather one-dimensional top-down approach of NATO's past. Future studies are needed to evaluate its implementation and performance.

Our interviews with key officials and close reading of alliance documents also provide further insights into how NATO is likely to interact with EDTs in the future. We identified NATO's approach to managing technology based on the four functions of generating, enabling, networking, and regulating. Likewise, we noted many potential political and technical bumps in the road. These include issues related to adoption challenges, intra-alliance capability gaps, protection of critical resources and human capital, and divisions over how to approach relations with China.

Russia's 2022 invasion of Ukraine revealed deep bonds of unification among NATO's member countries. The alliance has a history and structural base that should enable it to adapt to the age of EDTs, but we have highlighted significant obstacles standing in the way of NATO immediately reaping the benefits of such innovations. In the coming years, broad questions of EDT export controls and arms control are also likely to gain further prominence.[103] These will be difficult conversations, given the challenges posed by U.S.-China-Russia great-power competition and formulating anticipatory bans on emerging military technologies.[104] One thing is clear: to fully embrace the new era of defense innovation, NATO will have to innovate itself.

## NOTES

1. Katja Grace et al., "When Will AI Exceed Human Performance? Evidence from AI Experts," *Journal of Artificial Intelligence Research* 62 (July 2018), pp. 729–54.
2. Michiel van Amerongen, "Quantum Technologies in Defence & Security," *NATO Review*, 3 June 2021, www.nato.int/; Tristan A. Volpe, "Dual-Use Distinguishability: How 3D-Printing Shapes the Security Dilemma for Nuclear Programs," in "Emerging Technologies and Strategic Stability," ed. Todd S. Sechser, Neil Narang, and Caitlin Talmadge, special issue, *Journal of Strategic Studies* 42, no. 6 (August 2019), pp. 814–40.
3. James Johnson, "Artificial Intelligence & Future Warfare: Implications for International Security," *Defense & Security Analysis* 35, no. 2 (April 2019), pp. 150–52.
4. Olivier Schmitt, "Wartime Paradigms and the Future of Western Military Power," *International Affairs* 96, no. 2 (March 2020), pp. 401–18.
5. Rebecca Davis Gibbons et al., *The Altered Nuclear Order in the Wake of the Russia-Ukraine War* (Cambridge, MA: American Academy of Arts & Sciences, 2023), available at amacad.org/; Shannon Bugos, "Pentagon Sees Faster Chinese Nuclear Expansion," *Arms Control Today* 51, no. 10 (December 2021), pp. 26–28.
6. "Putin: Leader in Artificial Intelligence Will Rule World," *AP News*, 1 September 2017, apnews.com/.
7. Anna Nadibaidze, "Great Power Identity in Russia's Position on Autonomous Weapons Systems," *Contemporary Security Policy* 43, no. 3 (July 2022), p. 411.
8. Vinod K. Aggarwal and Andrew Reddie, "Putting the Biden Administration's 'New Economic Statecraft' in Context," *Lawfare*, 21 August 2023, lawfaremedia.org/.
9. Xiangning Wu, "Technology, Power, and Uncontrolled Great Power Strategic Competition between China and the United States," *China International Strategy Review* 2, no. 1 (June 2020), p. 100; Elsa B. Kania, "Testimony before the U.S.-China Economic and Security Review Commission Hearing on Trade, Technology, and Military-Civil Fusion: Chinese Military Innovation in Artificial Intelligence," *Center for a New American Security*, 7 June 2019, cnas.org/.
10. Stanford University Human-Centered Artificial Intelligence, *Artificial Intelligence Index Report 2023* (Stanford, CA: 2023), aiindex.stanford.edu/.
11. Shannon Bugos and Michael Klare, "Pentagon: Chinese Nuclear Arsenal Exceeds 400 Warheads," *Arms Control Today* 53, no. 1 (January/February 2023), p. 26.
12. Henrik Larsen, "Adapting NATO to Great-Power Competition," *Washington Quarterly* 45, no. 4 (Winter 2023), pp. 7–26.
13. NATO, *NATO 2022 Strategic Concept* (Madrid: 2022), www.nato.int/.
14. Ibid., para. 8.
15. Ibid., para. 13.
16. Ibid., para. 24.
17. Alexander K. Bollfrass and Stephen Herzog, "The War in Ukraine and Global Nuclear Order," *Survival* 64, no. 4 (August–September 2022), p. 7.
18. Dmitry Stefanovich, "Artificial Intelligence Advances in Russian Strategic Weapons," in *The Impact of Artificial Intelligence on Strategic Stability and Nuclear Risk*, vol. 3, *South*

43. NATO senior official no. 2, interview with authors, Brussels, March 2023.
44. Anna Nadibaidze and Nicolò Miotto, "The Impact of AI on Strategic Stability Is What States Make of It: Comparing US and Russian Discourses," *Journal for Peace and Nuclear Disarmament* 6, no. 1 (2023), pp. 47–67; Jeffrey Ding, *Deciphering China's AI Dream: The Context, Components, Capabilities, and Consequences of China's Strategy to Lead the World in AI* (Oxford, U.K.: Governance of AI Program, 2018), www.fhi.ox.ac.uk/.
45. NATO, "Brussels Summit Communiqué," press release no. (2021) 086, 14 June 2021, para. 3, www.nato.int/.
46. Ibid., para. 36.
47. Celeste A. Wallander, "Institutional Assets and Adaptability: NATO after the Cold War," *International Organization* 54, no. 4 (Autumn 2000), pp. 705–35.
48. Todd Sandler and Keith Hartley, "Economics of Alliances: The Lessons for Collective Action," *Journal for Economic Research* 39, no. 3 (September 2001), pp. 879–80.
49. NATO senior official no. 2 interview.
50. Adam Grissom, "The Future of Military Innovation Studies," *Journal of Strategic Studies* 29, no. 5 (October 2006), p. 907.
51. Michael C. Horowitz and Shira Pindyck, "What Is a Military Innovation and Why It Matters," *Journal of Strategic Studies* 46, no. 1 (February 2023), pp. 85–114.
52. Jeffrey A. Isaacson, Christopher Layne, and John Arquilla, *Predicting Military Innovation* (Santa Monica, CA: RAND, 1999), rand.org/.
53. Jon Schmid, Matthew Brummer, and Mark Zachary Taylor, "Innovation and Alliances," *Review of Policy Research* 34, no. 5 (September 2017), pp. 588–616; Matthew Brummer, "Innovation and Threats," *Defence and Peace Economics* 33, no. 5 (August 2022), pp. 563–84.
54. NATO senior official no. 2 interview.
55. Keith Hartley, "Defence Industrial Policy in a Military Alliance," in "Special Issue on Alliances," ed. Christopher Sprecher and Volker Krause, special issue, *Journal of Peace Research* 43, no. 4 (July 2006), pp. 473–89.
56. NATO senior official no. 3, interview with authors, Brussels, March 2023.
57. Office of Technology Assessment, *Arming Our Allies: Cooperation and Competition in Defense Technology* (Washington, DC: U.S. Government Printing Office, 1990), p. 3, available at ota.fas.org/.
58. Damon Coletta, "NATO-Russia Technical Cooperation: Unheralded Prospects," in *NATO's Return to Europe: Engaging Ukraine, Russia, and Beyond*, ed. Rebecca R. Moore and Damon Coletta (Washington, DC: Georgetown Univ. Press, 2017), p. 194.
59. NATO senior official no. 3 interview.
60. U.S. Defense Dept., *Contract Finance Study Report* (Washington, DC: Office of the Under Secretary of Defense for Acquisition and Sustainment Defense Pricing and Contracting, 2023), p. 35, acq.osd.mil/.
61. Jane Vaynman and Tristan A. Volpe, "Dual Use Deception: How Technology Shapes Cooperation in International Relations," *International Organization* 77, no. 3 (Summer 2023), pp. 599–632.
62. Jamie Shea, *Keeping NATO Relevant*, Policy Outlook (Washington, DC: Carnegie Endowment for International Peace, April 2012), p. 13, carnegieendowment.org/.
63. NATO, "Warsaw Summit Communiqué," press release no. (2016) 100, 9 July 2016, paras. 70, 76, www.nato.int/; NATO, "London Declaration," press release no. (2019) 115, 4 December 2019, para. 6, www.nato.int/; "Alliance Ground Surveillance (AGS)," *NATO*, 4 September 2023, www.nato.int/.
64. NATO, *Active Engagement, Modern Defence: Strategic Concept for the Defence and Security of the Members of the North Atlantic Treaty Organization* (Lisbon: 2010), para. 19, www.nato.int/.
65. NATO, *NATO 2022 Strategic Concept*, paras. 8, 13, 24.
66. NATO senior official no. 1 interview.
67. NATO, *Science & Technology Trends 2020–2040: Exploring the S&T Edge* (Brussels: NATO Science & Technology Organization, 2020), p. vii, www.nato.int/; NATO, *Science & Technology Trends 2023–2043: Across the Physical, Biological, and Information Domains*, vol. 1, *Overview* (Brussels: NATO Science & Technology Organization, 2023), pp. vi–vii, www.nato.int/.

68. NATO senior official no. 1 interview; "Emerging and Disruptive Technologies," *NATO*, 22 June 2023, www.nato.int/.
69. NATO senior official no. 4, interview with authors, Brussels, March 2023.
70. "Emerging and Disruptive Technologies."
71. NATO senior official no. 5, interview with authors, Brussels, March 2023.
72. NATO senior official no. 6, interview with authors, Brussels, March 2023.
73. Gary E. Marchant and Brad Allenby, "Soft Law: New Tools for Governing Emerging Technologies," *Bulletin of the Atomic Scientists* 73, no. 2 (March–April 2017), pp. 108–14.
74. NATO, *The Secretary General's Annual Report 2022* (Brussels: 2023), p. 78, www.nato.int/.
75. NATO senior official no. 6 interview.
76. NATO, *The Secretary General's Annual Report 2022*, p. 71.
77. NATO senior official no. 4 interview.
78. North Atlantic Treaty art. 2, 4 April 1949, available at www.nato.int/.
79. Sanne Verschuren "China's Hypersonic Weapons Tests Don't Have to Be a Sputnik Moment," *War on the Rocks*, 29 October 2021, warontherocks.com/.
80. Rose Gottemoeller et al., "Engaging with Emerged and Emerging Domains: Cyber, Space, and Technology in the 2022 NATO Strategic Concept," *Defence Studies* 22, no. 3 (July 2022), p. 522.
81. NATO senior official no. 4 interview.
82. "Budget," *DARPA*, darpa.mil/; "Emerging and Disruptive Technologies"; NATO senior official no. 5 interview; Vivienne Machi, "Are You Up for a Challenge? NATO Invites Startups to Compete for Cash," *Defense News*, 16 June 2023, defensenews.com/.
83. "Hub for EU Defence Innovation Established within EDA," *European Defence Agency*, 17 May 2022, eda.europa.eu/.
84. NATO senior official no. 6 interview.
85. Dominika Kunertova, "Drones Have Boots: Learning from Russia's War in Ukraine," *Contemporary Security Policy* 44, no. 4 (2023), pp. 576–91.
86. NATO, *Science & Technology Trends 2023–2043*.
87. NATO senior official no. 7, interview with authors, Brussels, March 2023.
88. NATO senior official no. 1 interview.
89. Robert Work, "The Third U.S. Offset Strategy and Its Implications for Partners and Allies" (remarks in Washington, DC, 28 January 2015), available at defense.gov/; Ian Livingston, "Technology and the 'Third Offset' Foster Innovation for the Force of the Future," *Brookings*, 9 December 2016, brookings.edu/.
90. Daniel Fiott, "A Revolution Too Far? US Defence Innovation, Europe and NATO's Military-Technological Gap," *Journal of Strategic Studies* 40, no. 3 (April 2017), pp. 417–37.
91. U.S. Defense Dept., *2023 National Defense Science and Technology Strategy* (Washington, DC: 2023), media.defense.gov/.
92. Office of Technology Assessment, *Arming Our Allies*, p. 3.
93. "Europe Can't Decide How to Unplug from China," *The Economist*, 15 May 2023, economist.com/.
94. NATO senior official no. 5 interview.
95. Barry R. Posen, *The Sources of Military Doctrine: France, Britain, and Germany between the World Wars* (Ithaca, NY: Cornell Univ. Press, 1984), p. 29.
96. For a discussion of the NATO Human Capital Innovation Policy see, for example, NATO, *Annual Report 2021* (Brussels: NATO Advisory Group on Emerging and Disruptive Technologies, 2021), pp. 24–25, www.nato.int/.
97. Eric Lofgren, Whitney M. McNamara, and Peter Modigliani, *Commission on Defense Innovation Adoption Interim Report* (Washington, DC: Atlantic Council, 2023), atlantic council.org/.
98. Fen Osler Hampson, "NATO's Conventional Doctrine: The Limits of Technological Improvement," *International Journal* 41, no. 1 (Winter 1985–86), p. 160.
99. John A. Burgess, "Emerging Technologies and the Security of Western Europe," in *Securing Europe's Future*, ed. Stephen J. Flanagan and Fen Osler Hampson (London: Routledge, 1986), pp. 65–84.

100. Dominika Kunertova and Stephen Herzog, "Emerging and Disruptive Technologies Transform, but Do Not Lift, the Fog of War: Evidence from Russia's War on Ukraine" (Helsinki: Finnish National Defence Univ., 2024); Kendrick Kuo, "Dangerous Changes: When Military Innovation Harms Combat Effectiveness," *International Security* 47, no. 2 (Fall 2022), pp. 48–87; Kathryn Hedgecock et al., "Emerging Technology and Strategy," *Defence Studies* 24, no. 1 (2024), pp. 133–40.
101. "Summary of the NATO Artificial Intelligence Strategy," *NATO*, 22 October 2021, www.nato.int/.
102. NATO senior official no. 8, interview with authors, Brussels, March 2023.
103. Justin Key Canfil and Elsa B. Kania, "Mapping State Participation in Military AI Governance Discussions," in *The Oxford Handbook of AI Governance*, ed. Justin B. Bullock et al. (New York: Oxford Univ. Press, 2024), pp. 895–913.
104. David M. Allison and Stephen Herzog, "'What about China?' and the Threat to US-Russian Nuclear Arms Control," *Bulletin of the Atomic Scientists* 76, no. 4 (July 2020), pp. 200–205; Justin Key Canfil, "Controlling Tomorrow: Explaining Anticipatory Arms Control for Emerging Military Technologies" (unpublished working paper, 24 April 2024), available at papers.ssrn.com/.